\documentclass{svmult}

\usepackage{makeidx}     
\usepackage{graphicx}    
\usepackage{multicol}    

\makeindex             

\newcommand{\dbar}{d \hspace*{-1.55ex}\raisebox{.7ex}{-}}

\begin{document}

\title*{Three-Nucleon Force in the $^4\rm{He}$ Scattering System}
\author{Hartmut M. Hofmann\inst{1}\and
Gerald M. Hale\inst{2}}
\institute{Institut f\"ur Theoretische Physik III, Universit\"at Erlangen--N\"urnberg, D--91058 Erlangen, Germany
\textit{hmh@theorie3.physik.uni-erlangen.de}
\and Theoretical Division Los Alamos National Laboratory, Los Alamos N.M. 87544, USA \textit{ ghale@lanl.gov}}
\maketitle

\begin{abstract}
We report on a consistent, microscopic calculation of the bound and
scattering states in the ${\rm ^4He}$ system employing modern realistic
two-nucleon and three-nucleon potentials
in the framework of the resonating group model (RGM). We present for
comparison with these microscopic RGM calculations the results from a
charge-independent, Coulomb-corrected $R$-matrix analysis of all types of
data for reactions in the $A=4$ system. Comparisons are made for selected
examples of
phase shifts and measurements from reactions sensitive to three-nucleon force
effects.
\end{abstract}

\section*{Introduction}

The $^4{\rm He}$ atomic nucleus is one of the best studied
few-body systems, both experimentally and theoretically, as
summarized in the recent $ A = 4$ compilation \cite {We1}.
Besides the many textbook examples of gross structure,
there are subtle points yielding large effects that are
only qualitatively understood. Except for \cite{hmgm} none of the existing
calculations aims at a complete understanding of the many
features of $^4{\rm He}$, which is not surprising in view of
the number of different phenomena studied so far \cite {We1}.
With the recent compilation \cite {We1} and the comprehensive
$R$-matrix analysis \cite {Ha1} of a large amount of
scattering data below $E_x=30$ MeV, a new, microscopic
calculation for the $^4{\rm He}$ system in this energy range using modern
realistic two- and three-nucleon forces 
is most desirable.

It is well known that realistic nucleon-nucleon ($NN$) forces cannot
reproduce the $\rm ^3H$, $\rm ^3$He, and $\rm ^4$He 
 binding energies. Three-nucleon
interactions (TNIs) are added to give the necessary small corrections
but they still fail to reproduce certain properties of the three
nucleon system, most notably the $A_{\rm y}$ analyzing power in $Nd$
scattering \cite{AY}.  Yet the 30\% deviation of $A_{\rm y}$ can be
resolved by tiny changes in the $Nd$ scattering phase shifts (on the
order of $0.1$ degrees \cite{PD3MEV, KIEVSKYND, ndpsa}).
Furthermore very many operators can contribute to a TNI and the lack
of stringent conditions in the three-nucleon system on the structure
of the TNI makes its application to other systems desirable.  In
\cite{hmgm} it was shown that although a realistic $NN$ force can generally
reproduce the $^4$He system, there remain differences,
most notably in the analyzing powers.  Since the intensely studied $\rm
^4He$ system \cite{We1} is unfortunately very difficult to 
describe due to the many resonances and the $\rm^4He$ bound state,
the much simpler systems $p-^3$He and
$n-^3$H where data exist in the energy range of interest were investigated
in \cite{bp}.

The essential findings of this work are
that realistic $NN$ interactions describe most of the phase shifts
quite well but fail to reproduce the $^3P_2$ and $^3P_0$ phase shifts.
The calculated splitting between these two channels is much too small, and
neither the Urbana IX (UIX) \cite{U9} nor the TLA \cite{TLA}
 three-nucleon force is able to improve
the splitting significantly.  In fact, there it is more important to include in
the calculation  negative parity states of the three-nucleon subsystem
than one of these two TNIs. These findings suggest that new contributions to
the $NNN$ force acting on the $P$-waves should be considered, like an
$LS$ type TNI, as proposed in \cite{KIEVSKYLS} for the $N-d$ analyzing
powers, or the $V_3^*$ operators proposed in \cite{V3SCHADOW}.
Based on these findings we choose as $NN$ force only the AV18 \cite{av18}
and as TNI the Urbana IX \cite{U9} and in addition the $V_3^*$  \cite{V3SCHADOW}. 

We organize the paper in the following way. The next section contains a brief
discussion of the Resonating Group Model calculation together with the model
spaces used. Then we compare R-matrix and RGM results of a few typical
examples of scattering phase shifts for various model spaces and combinations
of interactions. Finally we compare with data for examples sensitive to the
TNIs.

\section*{RGM and model space}

We use the Resonating Group Model \cite{RRGM, VIEWEG, TANG}
to compute the scattering in the $\rm ^4$He system  
using the Kohn-Hulth\'en variational principle \cite{KOHN}. The main
technical problem is the evaluation of the many-body matrix elements
in coordinate space. The restriction to a Gaussian basis for the
radial dependencies of the wave function allows for a fast and
efficient calculation of the individual matrix elements \cite{RRGM,
  TANG}. However, to use these techniques the potentials must  
also be given in terms of Gaussians. In this work we use suitably
parametrized versions of the Argonne AV18 \cite{av18} 
 $NN$ potential and the Urbana IX \cite{U9}
and  the $V_3^*$  \cite{V3SCHADOW}. The inclusion of an additional TNI requires almost two orders of magnitude more 
computing power than the realistic $NN$ forces alone.

In the $^4$He system we use a model space with six two-fragment
channels, namely the $p - ^3$H, the $n - ^3$He, the $^2$H$ - ^2$H,
the singlet deuteron and deuteron \dbar - $^2$H, the \dbar - \dbar, and the
$(nn) - (pp)$ channels. The last three are an approximation to 
the three- and four-body breakup channels that cannot in practice   
be treated within the RGM. The $^4$He is treated as four clusters
in the framework of the RGM to allow for the required internal orbital
angular momenta of $\rm ^3$He, $^3\rm{H}$  or $\rm ^2$H.

For the scattering calculation we include the $S$, $P$ and 
$D$ wave contributions to the $J^\pi = 0^+, 1^+, 2^+, 0^-, 1^- $  and 
$2^-$ channels. From the $R$-matrix analysis these channels are known  to give
 essentially the experimental data. (We discuss cases where this is not the case.)
  The full wave functions for these
channels contain over 200 different spin and orbital angular momentum
configurations, hence it is too complicated to be given in detail.
The simplest wave functions we use for $^3$He are those described in \cite{bp}.

This small 29-dimensional model space yields -6.37 MeV binding energy,
an {\em rms} radius of 1.78 fm and a $D$ state probability of 7.7\%  
for the $\rm ^3$He using AV18. In order to avoid fake effects
the relative thresholds in $^4$He should be reproduced well, therefore we
used also a 35-dimensional modelspace, called large,by allowing additional
configurations with two orbital angular momenta on the two Jacoby coordinates
 yielding -6.69 MeV binding energy.
This must be compared to
$-6.92$MeV known from Faddeev calculations \cite{FAD}.
For the deuteron we use the structure given in \cite{hmgm}, yielding a binding
energy of -1.921 MeV, which could be easily improved. But then the relative
threshold energies deteriorate, see table \ref{thres}.
All the Gaussian width parameters were obtained by a non-linear
optimization using a genetic algorithm \cite{CWGEN} for the combination AV18 and
UIX.

Once the
fragment wave functions are fixed the scattering problem is solved
with our RGM code relying on the Kohn-Hulth\'en variational principle
\cite{KOHN}:
$$\delta \big( \langle \Psi_t | H-E | \Psi_t \rangle -
\frac{1}{2}a_{ll}\big) =0,
$$
where $a_{ij}$ denotes the reactance matrix.

The model space described above (consisting of four to ten physical   
scattering channels for each $J^\pi$) is by no means sufficient to
find reasonable results. So-called distortion or pseudo-inelastic
channels \cite{TANG} have to be added to improve the description
of the wave function within the interaction region. Accordingly, the 
distortion channels have no asymptotic part.

For practical purposes it is obvious to reuse some of the already
calculated matrix elements as additional distortion channels. In that
way we include all the positive parity states of the three-nucleon
subsystems with $J^\pi_3 \le 5/2^+$ in our calculation. However, it was
recently pointed out by A.\ Fonseca \cite{NT} that states
having a negative parity $J_3^-$ in the three-nucleon fragment
increase the $n -^3$H cross section noteably. Therefore we also   
added the appropriate distortion channels in a similar complexity as 
in the $J_3^+$ case to our calculation, thereby roughly doubling the size of
the model space. The 20 percent increase of the model space for the 3N bound states
from 29 to 35 resulted in almost a factor of two in the computing time.
The parameters of the $V_3^*$ TNI were adjusted that the binding energy of
triton and $^3$He did change by less than 10 keV. Therefore we do not give the
corresponding energies in table \ref{thres}.

\begin{table}
\centering
 \caption{\label{thres} Comparison of experimental and
calculated total binding energies and relative thresholds (in MeV) for 
the various potential models used}
\vskip 0.2cm
\begin{tabular}{c|c|c|c|c}
\hline\noalign{\smallskip}
potential & \multicolumn{2}{c}{$E_{bin}$} & \multicolumn{2}{c}{$E_{thres}$} \\
    & $^3$H & $^3$He & $^3{\rm He}-p$  &  $ { d - d} $ \\
\noalign{\smallskip}\hline\noalign{\smallskip}
av18    & -7.068& -6.370 & 0.698 & 3.227 \\
av18, large  & -7.413& -6.588 & 0.725 & 3.572 \\
av18 + UIX & -7.586 & -6.875 & 0.710 & 3.745 \\
av18 + UIX,large & -8.241 & -7.493 & 0.748 & 4.400 \\
exp.    & -8.481 & -7.718 & 0.763 & 4.033\\
\noalign{\smallskip}\hline
\end{tabular}
\end{table}

\section*{Results}
Since we are mainly interested in the effects of 3N-forces, we mention the bound
state results only briefly. In the large model space AV18 plus UIX yields
-27.81 MeV close to the experimental -28.296 MeV, to which the parameters of
the UIX are fitted to. Although the parameters of the $V_3^*$ TNI were chosen
as to give only minute changes in the three-nucleon system, for $^4$He it
resulted in 650 keV additional binding.

\begin{figure}
\centering
\includegraphics[height=8cm]{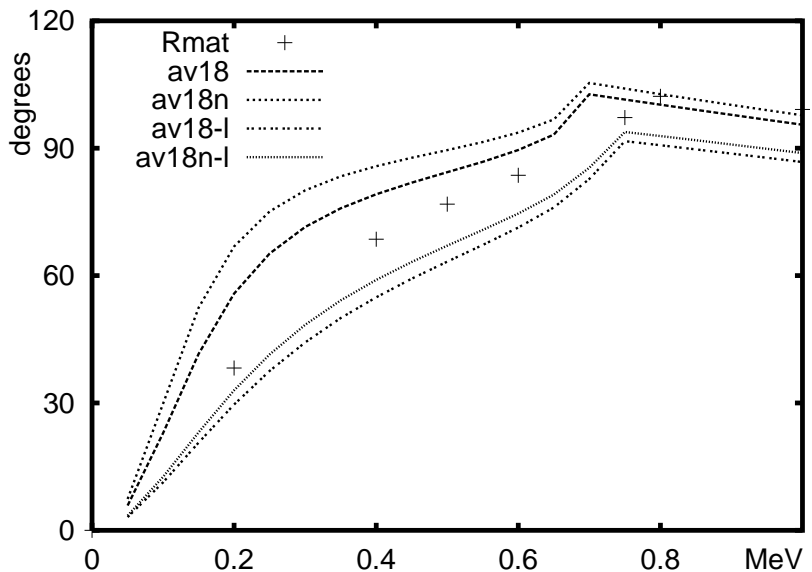}
\caption{Comparison of the $0^+$ triton-proton phase shifts from the R-matrix
analysis (crosses) and calculations employing AV18 in the small model space
(av18), adding negative parity distortion channels (av18n), for the large model
space (av18-l) and adding negative parity distortion channels (av18n-l)}
\label{nullp}
\end{figure}

\begin{figure}
\centering
\includegraphics[height=8cm]{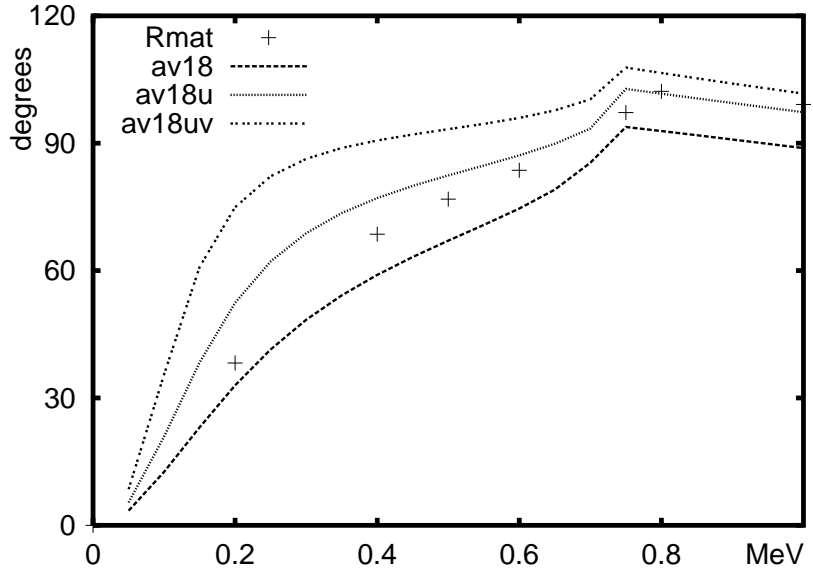}
\caption{As fig \ref{nullp}, but R-matrix results (crosses) are compared to the 
full NN-calculation (av18), adding UIX (av18u) and adding $V_3^*$ (av18uv)}
\label{0-NNN}
\end{figure}

\begin{figure}
\centering
\includegraphics[height=8cm]{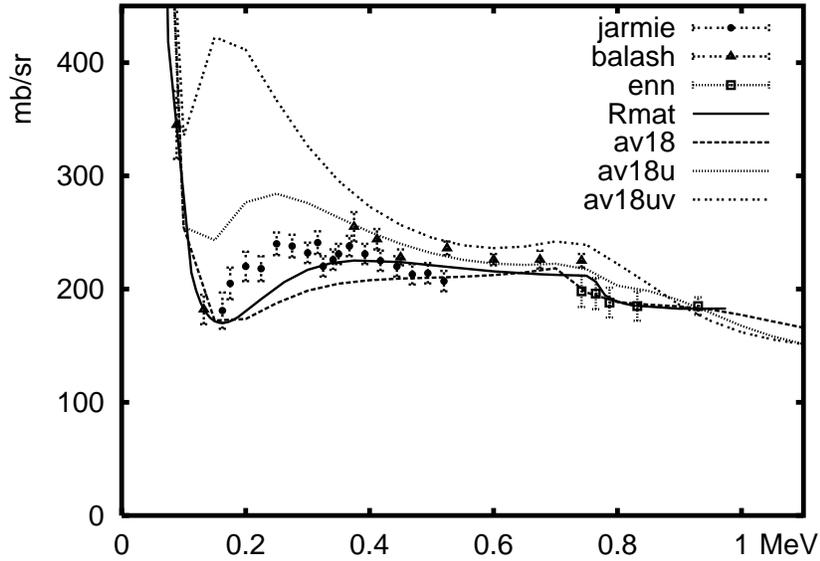}
\caption{Differential triton-proton cross section at 120 degrees as function
of energy. The data are from Jarmie \cite{jarmie}, Balashko \cite{bash}, and
Ennis \cite{enn}. The other lines are as in fig \ref{0-NNN}.}
\label{120_NNN}
\end{figure}

The most detailed comparison between calculation and data is on the level of
an energy
dependent phase-shift analysis. This is given by the R-matrix analysis as 
described in \cite{hmgm} in detail. The lowest channel, triton-proton contains
the intriguing first exited state of $^4$He, a $0^+$-resonance, sometimes
considered a breathing mode, which is clearly seen
in the $0^+$ phase shift, see fig. \ref{nullp}, but does not 
show-up in the angular distributions, see fig. \ref{120_NNN}.
Neglecting the Coulomb force this resonance is moved just below the p-triton
threshold, i.e. it becomes a bound state, bound by less than 50 keV depending
on the force and model space used. Therefore all approaches, which neglect
the Coulomb force like \cite{NT} cannot aim at this energy region.
In fig. \ref{nullp} the
R-matrix results are compared to the pure NN-calculations for various model
spaces. For the small model space the calculation is slightly above the
R-matrix data. Adding the negative parity distortion channels, we find a small increase
due to the enlarged attraction. This effect is much smaller than the one 
found in
triton-neutron scattering \cite{NT,bp}. Increasing the 3N-model space
reduces the phase shifts considerably, due to the better $^4$He binding
and the additional thresholds shifted to higher energy, see table \ref{thres}.
In fig. \ref{0-NNN} we find
strong sensitivity to the additional TNI. This sensitivity in specific partial
waves might help to unravel the operator structure of the TNI, especially as all
the thresholds are unchanged from UIX to $V_3^*$.

For 58 and 120 degrees exist measured exitation functions for the p-triton
differential cross section. For the forward
angle the R-matrix analysis is on top of the data \cite{jarmie}, the pure
NN-calculation a bit below, with UIX almost on top of the data and
with $V_3^*$ a bit above. Since there are only minor differencies, we do not
show them. For the backward angle,however, see fig. \ref{120_NNN}, even the
 R-matrix analysis
cannot fully reproduce the data in the Coulomb-nuclear interference region.
The data from \cite{bash} seem to be consistently above those from \cite{jarmie}
and \cite{enn}. In the Coulomb-nuclear interference the sensitivity to
TNI is very large, whereas around the (p,n) threshold data, R-matrix and 
microscopic calculations agree essentially. Since all the data are very old
and not consistent a new measurement is urgently 
called for. Unfortunately due to
radiation hazards of the triton this is not very likely.

A recent measurement of the real parts of the neutron-$^3$He spin-dependent
scattering lengths \cite{epj} $a_0$ = 7.370(58) fm and $a_1$ = 3.278(53) fm
is therefore very important. The corresponding R-matrix results are
$a_0$ = 7.398 fm and $a_1$ = 3.257 fm. Due to the strong coupling via the
$0^+$ resonance $a_0$ has a large imaginary part.  Hence, the numerical extraction
of this scattering length is not easy. Therefore we give only preliminary
numbers in the following. For the large model space the microscopic
calculations yield $a_0 = 7.402 \div 7.590$ fm and $a_1 = 3.289 \div 3.424$ fm,
depending on the combination of forces.
These results are close enough to be used for a further determination of the
operator structure of the three-nucleon force.

\begin{figure}
\centering
\includegraphics[height=8cm]{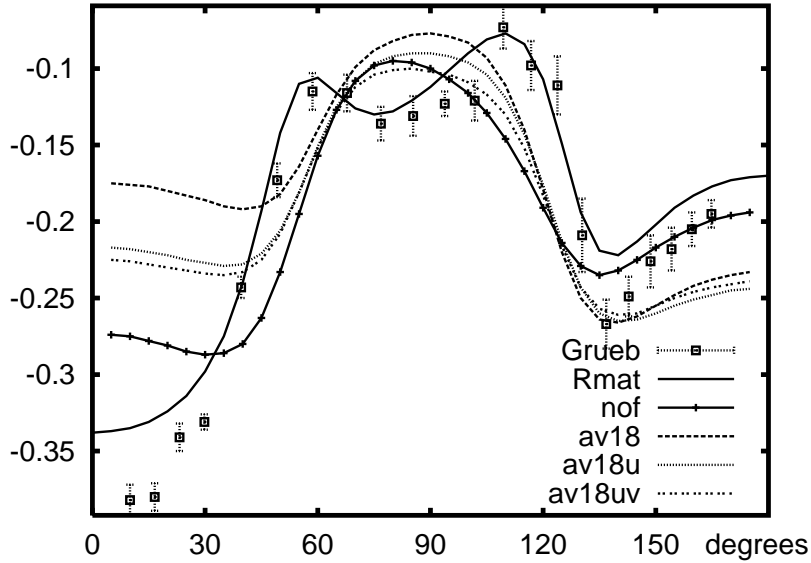}
\caption{Tensor analysing power $T_{20}$ for the reaction $^2$H(d,p)$^3$H at
2.0 MeV center-of-mass energy. The data are from \cite{grueb}, the lines are as in fig \ref{0-NNN} and additionally the R-matrix results without F-waves (nof).}
\label{t20_NNN}
\end{figure}

Out of the many possible data we choose an example, which demonstrates the
limitation of our total partial wave model space. The well studied 
deuteron-deuteron reactions allow a detailed comparison since together with differential
cross sections for many energies also analysing powers are available. The
tensor analysing power $T_{20}$
displays a pronounced angular distribution, see fig. \ref{t20_NNN}. The full
R-matrix analysis nicely reproduces the data, whereas all microscopic
calculations fail at forward angles and show no sign of the double-hump
structure. Omitting all F-wave S-matrix elements in the R-matrix analysis
corresponding to the model space used for the RGM calculation yields an angular
dependence quite similar to the microscopic results, see fig. \ref{t20_NNN}.
This demonstrates the
need for $L_{rel} = 3$ partial waves in the RGM calculations.

\section*{Conclusion}
We presented a complete microscopic calculation in the $^4$He system
employing modern realistic two- and three-nucleon forces. We demonstrated
that in specific examples 
the inclusion of NNN-forces yields large effects in phase shifts,
differential cross sections and analysing powers.
Hence, the $^4$He-system seems
well suited for a detailed study of different NNN-forces, especially since
a comprehensive R-matrix analysis exists, which reproduces the vast amount
of data for various reactions very well, thus allowing for a comparison on
the level of individual partial waves. Therefore a determination of the
operator structure of the NNN-force is within reach, provided the microscopic
calculations are converged. To aim at this goal the internal triton and 
$^3$He wave functions have to be improved, such that the binding energy is 
within say 50 keV of the experimental
value. This can only be achieved by
 a major increase of the model space. For the deuteron the
corresponding modification is trivial. To  describe the deuteron-deuteron
reactions, we have to allow also for F-waves on all relative coordinates.
All these improvements are relative straightforward, but require about a factor
5 in the CPU-time compared to the needs of the work reported here.

Acknowledments: The work of H.M.H is supported by the BMBF (contract 06ER926) and
that of G.M.H. by
the Department of Energy. The grant of computer time at the HLRB and the
RRZE is gratefully
acknowledged. We want to thank G. Wellein and G. Hager at the RRZE for their
help.


\printindex
\end{document}